# Strain Healing of Spin-Orbit Coupling: A Cause for Enhanced Magnetic Moment in Epitaxial SrRuO$_3$ Thin Films

Shekhar Tyagi, Gaurav Sharma, D. M. Phase, V. R. Reddy and V. G. Sathe

UGC-DAE Consortium for Scientific Research, University Campus, Khandwa Road, Indore-452001, India

**Abstract**

Enhanced magnetic moment and coercivity in SrRuO$_3$ thin films are significant issues for advanced technological usages and hence are researched extensively in recent times. Most of the previous reports on thin films with enhanced magnetic moment attributed the high spin state for the enhancement. Our magnetization results show high magnetic moment of 3.3 $\mu_B$/Ru ion in the epitaxial thin films grown on LSAT substrate against 1.2 $\mu_B$/Ru ion observed in bulk compound. Contrary to the expectation the Ru ions are found to be in low spin state and the orbital moment is shown to be contributing significantly in the enhancement of magnetic moment. We employed x-ray absorption spectroscopy and resonant valance band spectroscopy to probe the spin state and orbital contributions in these films. The existence of strong spin-orbit coupling responsible for the de-quenching of the 4d orbitals is confirmed by the observation of the non-statistical large branching ratio at the Ru M$_{2,3}$ absorption edges. The relaxation of orbital quenching by strain engineering provides a new tool for enhancing magnetic moment. Strain disorder is shown to be an efficient mean to control the spin-orbit coupling.

**1. Introduction:**

SrRuO$_3$ (SRO) [1] is getting wide attention from the applied scientific community for its unique applications like electrodes [2] or conducting layers in epitaxial hetero-structures of functional oxides and magnetic tunnel junctions [3,4,5] due to the robust spin polarization at the interface of the ferromagnetic oxide and the metal. Few materials are available in literature that has such variety of properties. SrRuO$_3$ shows itinerant ferromagnetism, unusual transport properties, Fermi liquid nature at low temperatures [6], bad metal behaviour at high temperatures [7], high-perpendicular remnant magnetization and large magneto-optical constant in thin films [8].

It is by now widely accepted that a small disorder in SrRuO$_3$ produces large variation in its physical properties. Previous reports suggest that the magnetic properties can be significantly modified by introduction of strain [9,10,11,12]. These studies showed that the physical properties of SRO are dissimilar in strained and strain relaxed films. For example; Q. Gan et al. [12] showed that the transition temperature reduces with enhancement in substrate induced strain in films of SRO grown on (001) oriented substrate. In fully strained films, the in-plane and out-of-plane inter atomic distances are considerably dissimilar, resulting in variation in the overlap of *Ru*: t$_{2g}$ and O: *2p* orbitals in the two directions. This significantly modifies the physical properties as the electronic conduction and magnetic properties are known to be associated with the degree of hybridization of the *Ru: 4d* orbitals and O: *2p* orbitals.

Recently, there is a race among researchers for attaining enhancement in magnetic moment in the films of SRO. Bulk SRO shows low spin state and a magnetic moment of ~1.2 $\mu_B$/Ru ion. Alexander et al. [13] showed enhanced magnetic moment i.e. 1.7$\mu_B$/Ru and 1.9$\mu_B$/Ru in SRO films grown on (001) SrTiO$_3$ and LSAT substrate, respectively. Importantly the saturation moment was found to be in-sensitive to the film thickness and the lattice distortions are shown to be responsible for controlling the magnetic ground state in SRO films. Murtaza et al. [14] reported a large saturation magnetic moment in the range of 2 - 2.4 $\mu_B$ in Ru deficient films which was attributed to the stabilization of the high spin $Ru^{+4}$ state. Alexander et al. [15] proposed that the high spin state of the Ru ions can be stablized through controlled lattice distortion introduced by the compressive nature of in-plane strain and the orientation of the film. It was stated that in high spin state, the orbital overlapping decreases and an enhancement in the density of states at the Fermi level is observed due to newly split $e_g$ band. It was proposed that the growth in the (111) orientation is responsible for the stabilization of the high spin state. In contrast to this report, Bowha et al. [16] and Agrestini et al. [17] negated the role of orientation of the film growth in stabilization of the high spin state. Bowha et al. [16] advocated that the nearest neighbour distances of the Ru ions decides the magnetic properties of the SRO thin films and found no signatures of the high-spin states in their study on the (111)-oriented SRO thin films. Agrestini et al. [17] argued that the stabilization of the high spin state with $S = 2$ would be too costly in energy and hence it is not possible to stabilize high spin state in $Ru^{4+}$. They proved that the thin films of SRO grown on (111) and (001) oriented $SrTiO_3$ substrates are in low spin states with quenched orbital moment. These authors ruled out the hypothesis of a compressive strain-induced spin state transition. However, recently Ning et al. [18] reported 4 $\mu_B$/Ru moment in the (111)-oriented SRO films grown on $SrTiO_3$ substrate. They observed 2 $\mu_B$/Ru and 3 $\mu_B$/Ru moment in the (001)- and (110)-oriented films, respectively. These magnetization values are attributed to the high-spin state, low-spin state and mixed (low and high) spin states of Ru ion in the (111)-, (001)-, and (110)-oriented SRO films, respectively. Hence, the origin of enhancement in the magnetic moment remains a controversial issue.

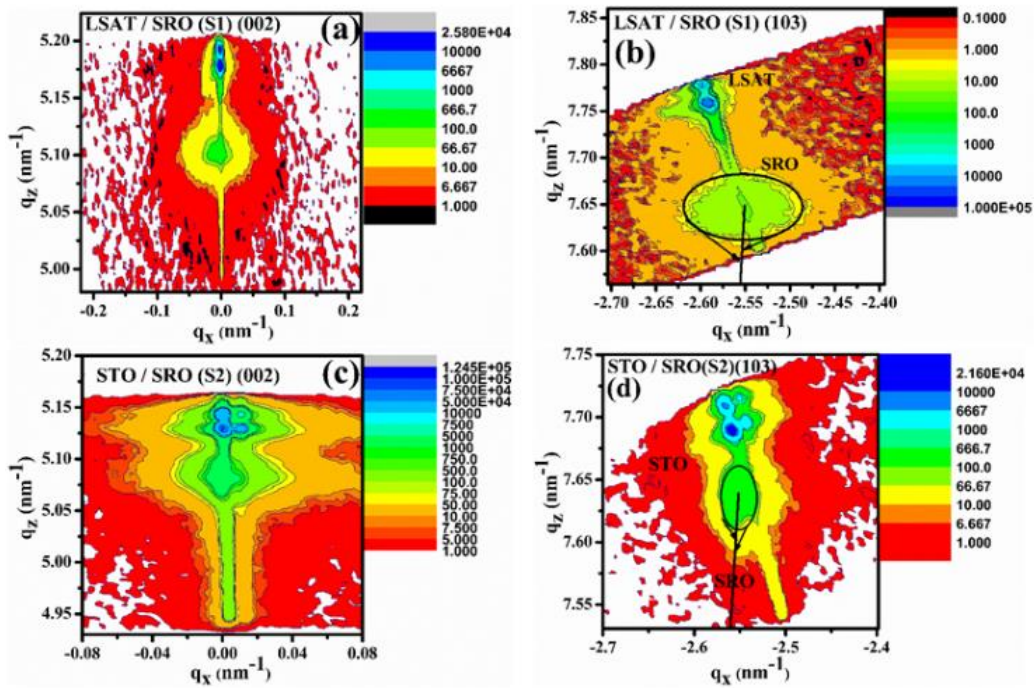

**Figure 1:-** The Reciprocal space mapping (RSM) plot recorded at the symmetric (002) and asymmetric (103) plane: (a) and (b) shows the RSM plot corresponding to the LSAT/SRO (S1) film while (c) and (d) shows the RSM plot corresponding to the STO/SRO (S2) epitaxial thin film. The circles illustrates the broadening of the reciprocal lattice point and arrows indicates amount of peak broadening perpendicular and parallel to the reflecting plane.

The enhancement in magnetic moment in SRO films can be due to two mechanisms; first is due to stabilization of the high spin state of the Ru ions with S=2 and the second possibility is due to orbital moment contribution. The second possibility is operative only when the orbital quenching is disrupted.

One of the ways, we thought to disrupt the quenching of orbital moment is to weaken the crystal field effect. This can be achieved by lowering the local symmetry of the oxygen octahedra surrounding the $Ru^{4+}$ ion by introduction of large disorder. Large disorder in films can be created by creating defect states but in that case the spin moment is also expected to decrease significantly. The other method is to introduce large strain disorder in films during growth, which will not affect the spin moment but is expected to affect the local crystal field. It has previously been reported that the orbital angular moment weakens due to reduction in hybridization of Ru4+ ions with the surrounding oxygens [19].

In order to address above issues and to achieve strain disorder, thin films of $SrRuO_3$ are grown on (001) oriented $SrTiO_3$ and LSAT substrates of various thicknesses providing different strain relaxation states. Intentionally, films with strain disorder were chosen for the investigation of their magnetic properties. Our study show that both the films remain in low spin state with different degrees of orbital quenching resulting in enhanced magnetic moment in the film grown on LSAT.

## 2. Experimental Techniques:

The $SrRuO_3$ is prepared by standard solid state reaction method and characterised by x-ray diffraction and Raman spectroscopic measurements. It is found single phase in nature and showed the ferromagnetic transition at 160K, exactly matching with literature [20,21,22]. A dense pellet of 20mm diameter was prepared to be used as a target during deposition of thin films of $SrRuO_3$ on (MTI-corp. USA, make) (001) oriented single crystal substrates of $SrTiO_3$ and LSAT, using pulsed laser deposition technique. A pulsed Excimer laser with 248nm wavelength (pulse width = 20 ns) providing an energy density of $3.6 Jcm^{-2}$ and repetition rate of 10Hz is used for the deposition. The substrates were kept at 750°C during deposition, in the presence of oxygen gas partial pressure of 0.130 mbar. After several test depositions the cooling rate after deposition was kept at 40°C per minute that resulted in best quality films. For confirming the phase purity x-ray diffraction was carried out using BRUKER D2 Phaser (Cu K- α radiation). The reciprocal space mapping measurements were carried out using Bruker D8-Discover high resolution X-ray diffractometer (HRXRD). Magnetic properties of the films was examined using 7-Tesla SQUID-vibrating sample magnetometer (SVSM; Quantum Design Inc., USA). Valence band spectroscopic measurements are performed at angle-integrated photoemission spectroscopy (AIPES) beamline on Indus-1 synchrotron radiation source at RRCAT, Indore, India. To measure the valance band spectra the base pressure in the experimental chamber was of the order of $10^{-10}$ Torr. Spectra were collected at 300 K with incident photon energy of 48 eV (off-resonance) and 52 eV (on-resonance). Fermi level was aligned by recording the valence band spectra of the Ag foil, which is also mounted along with the sample on the holder. Electronic properties of both the films and the bulk compound was analysed using X-ray absorption spectroscopy technique measured in total electron yield (TEY) mode across O(oxygen) K-edge at soft X-ray absorption spectroscopy (SXAS) beamline, BL-1[23] at Indus-2 RRCAT, Indore, India.

## 3. Results and Discussions:

In order to probe the effect of strain disorder, three samples were mainly investigated. Thin films of $SrRuO_3$ grown on LSAT substrate (S1), $SrTiO_3$ substrate (S2) and pellet of bulk $SrRuO_3$. The films grown using pulsed laser deposition technique were thoroughly characterized using x-ray diffraction technique used in various modes. The x-ray reciprocal space map analysis was carried out at the symmetric (002) and asymmetric (103) reflection which provided details of strain disorder and epitaxy. Figure 1 (a) and (b) shows the RSM carried out on S1 film while figure 1 (c) and (d) show RSM carried out on the S2 film along symmetric (002) and asymmetric (103) plane, respectively. The in-plane *"a"* and out-of-plane *"c"* lattice parameters deduced from the RSM analysis are a=3.926 Å, 3.920 Å; and c= 3.924 Å, 3.926 Å for the S1 and S2 films respectively. The strain values resulted from the lattice parameters in the in-plane direction are -0.07% and -0.22% while in the out-of-plane direction are -0.12% and -0.09% for the S1 and S2 films, respectively. The RSM study thus suggests that both the films are nearly relaxed with nominal strained values. The asymmetric reciprocal lattice point corresponding to the films showed significant broadening as indicated by a circle [Figure 1 (b), (d)]. The length of the tangential arrows with the circle represents the amount of peak broadening perpendicular and parallel to the reflecting plane. It is well known that the broadening perpendicular to the reflecting plane corresponds to the mosaic spread while the broadening parallel to the reflecting plane corresponds to the lateral correlation length; smaller the correlation length broader the point. From the RSM measurements the mosaic spread representing the angular range of the low-angle grain boundaries were calculated. The calculated angular values are $1.10^0$ and $0.22^0$ corresponding to the S1 and S2 films, respectively. We have also calculated the lateral correlation length that are found to be quite different in the two films. The calculated lateral correlation length is found to be 3.45 Å in S1 film and 15.44 Å in S2 film which suggests that the strain is getting modulated in the adjacent unit cell in S1 film while in S2 film, strain is modulating after four unit cells. Thus, it can be concluded that the local strain disorder is larger in S1 film than in S2 film.

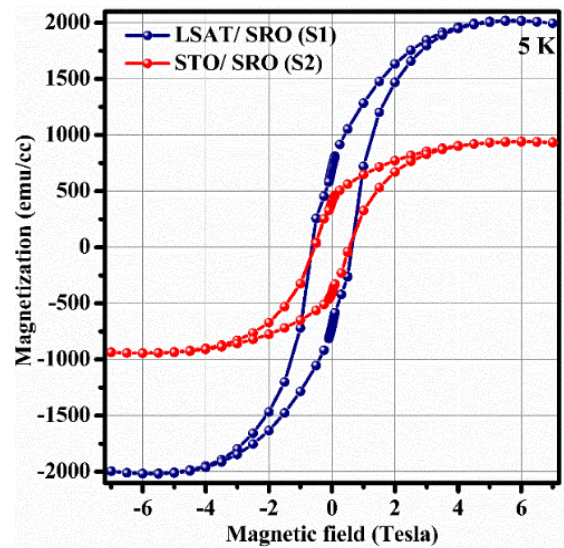

**Figure 2:-** The magnetization (M) as a function applied magnetic field (H) for LSAT/SRO (S1) and STO/SRO (S2) thin films measured at 5K.

The films were further characterized by Raman spectroscopy (not shown here). The Raman spectra (both character and position of the modes) matched with previously reported spectra for epitaxial thin films [24].

To find out the saturation magnetic moment, we have carried out the magnetization vs magnetic field (M-H) measurements at 5K for both the films (figure 2). The diamagnetic contribution arising from the substrate have been subtracted using standard procedure [25]. The saturated magnetic moment calculated from the M-H curves was found to be 3.3 $\mu_B$ / $Ru^{+4}$ for S1 film and 1.5 $\mu_B$ / $Ru^{+4}$ for S2 film. These values are significantly higher when compared to bulk. In the bulk SRO, a magnetic moment of ~ 1.2 $\mu_B/Ru^{4+}$ was reported that is attributed to low-spin state configuration due to the large crystal field splitting [15]. Importantly, the coercivity of the hysteresis loop was found to be of the order of 0.65 and 0.46 tesla in S1 and S2 film, respectively. Such a large value of the coercivity in S1 can be attributed to the higher local strain disorder as deduced from the RSM study, as it is likely to enhance the domain pinning.

In order to understand the origin of enhanced magnetic moment, possible effective magnetic moments taking different configurations were calculated that is given below.

**1.** $Ru^{4+}$ in high spin state and orbital moment is completely quenched

*l=0, s=2 and j=2, g = 2 giving effective magnetic moment = 4.8 $\mu_B$,*

**2.** $Ru^{4+}$ in low spin state and orbital moment is completely quenched

*l=0, s=1 and j=1, g = 1 giving effective magnetic moment = 1.4 $\mu_B$,*

**3.** $Ru^{4+}$ in low spin state and orbital moment is not quenched

*l=2, s=1 and j=3, g = 4/3 giving effective magnetic moment = 4.6 $\mu_B$*

where, g is Lande g-factor, l and s are orbital and spin angular moment, while, j is total angular moment.

These calculations suggests that the film S2 is in low spin state with orbital moment fully quenched as observed in bulk compound. On the other hand the experimentally observed enhanced magnetic moment in S1 cannot be fully explained by considering any of the above configurations.

In order to estimate the spin state of the $Ru^{4+}$ ions in these samples the oxygen K-edge x-ray absorption spectra (XAS) was recorded and is presented in figure 3. The resonant photoemission spectra (PES) recorded in the on-resonance

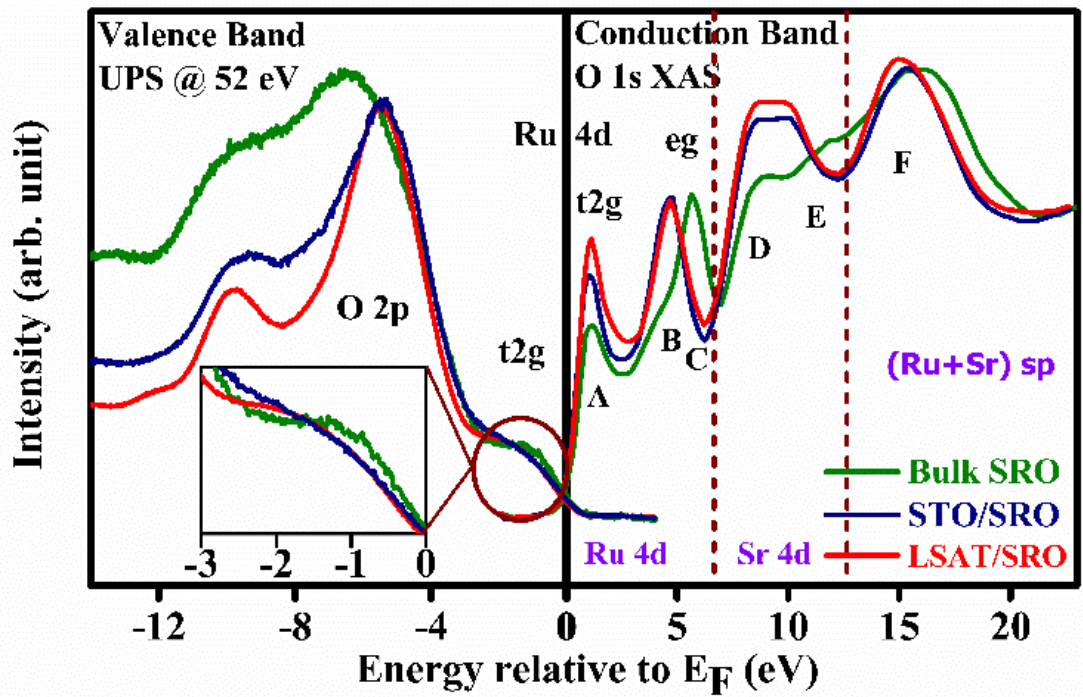

**Figure 3:-** The soft X-ray absorption spectroscopic spectra recorded on the LSAT/SRO (S1), STO/SRO (S2) and bulk SRO at the oxygen *1s* (K-edge) presented in the right panel. For better information the valance band spectra recorded at the "on" resonance condition as a function of relative energy to the fermi energy ($E_F$) is also presented in the left panel.

($h\nu$=52 eV) condition is also presented in the same figure for better understanding. The XAS provides information on unoccupied density of states while the PES provides information about the occupied density of states. The combination, thus, provides detailed electronic structure of the Ru ions. The XAS spectra characterises the transition from O *1s* core level to unoccupied O *2p* level in the conduction band. It is well known that these conduction band levels are strongly hybridized with Ru *4d* states. Thus, the O *1s* XAS spectra provides direct information about unoccupied Ru bands (*4d*) through the hybridization [26,27]. We carried out the analysis of the O *1s* XAS spectra of the bulk SRO compound by following the terminology used in refs. [22, 23,28]. Accordingly, the features (A, B and C) near the edge are attributed to the Ru *4d* bands in figure 3. The Ru *4d* band is known to get split into t2g and eg sub-bands due to the octahedral crystal-field effect. The eg band further splits due to the intra-atomic exchange into a majority and a minority band. The feature A corresponds to the localized t2g sub-band (π-bonding) while features B and C corresponds to the delocalized eg sub-band (σ-bonding). The features D and E correspond to the Sr *3d* bands, whereas the feature F are attributed to the metal (Ru + Sr) *sp* bands due to the (Ru + Sr) *sp*–O *2p* interactions.

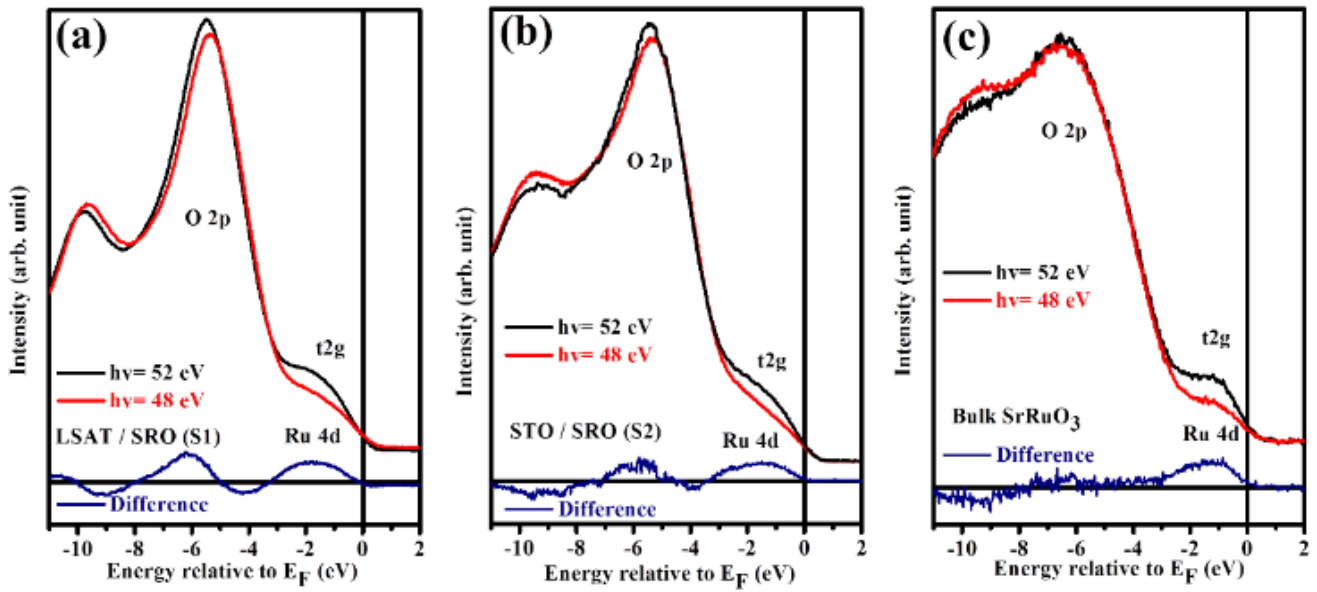

**Figure 4:-** Comparison of Valance band spectra (VBS) recorded in the "on" and "off" resonance condition is illustrated in (a), (b) and (c) corresponding to the LSAT/SRO (S1), STO/SRO (S2) thin films and bulk SRO, respectively. The lower curve represents the difference spectra obtained by subtracting the spectra recorded in on" and "off" resonance conditions.

Interestingly, the film S1 and S2 shows major differences in the O *1s* XAS spectra when compared to spectra shown by the bulk compound. It is noted that the position of the feature A is nearly same in bulk and films, however, the relative intensity of the A peak reflecting Ru t2g band show significant enhancement in films when compared to the bulk compound. This peak is relatively broad in S1 film than in S2 and bulk. The shape profile of B and C features reflecting Ru eg band show a drastic variation in films then in bulk. The two features merge completely in films forming a single band which show a shift towards lower binding energy without any change in the intensity.

These changes in the spectrum suggests that the degree of hybridization of the O *2p* states with Ru *4d* states is substantially different in the thin films than in bulk. The crystal field splitting (separation between t2g and eg bands) is significantly lower in films (3.56 eV for S1 film and 3.63 eV for S2 film) than in the bulk (4.58 eV). One of the plausible reasons for decrease in crystal field splitting can be reduced σ-bonding strength and enhanced π-bonding strength due to variation in octahedral distortion in the films. The mutual exchange of the strength between the bonding of orbitals decides the change in the crystal field splitting.

Compared to the bulk, the enhanced intensity of the feature A in the films indicates that the unoccupied density of state in Ru t2g band are getting enhanced in the films, particularly in S1. The change in shape profile of the Ru eg band indicates the modulation in the exchange effects between the majority and minority spin contributions [29]. The XAS spectra at higher binding energy of the bulk and films show significant change in spectral character which reveals modification in the hybridization state of the Sr *4d* bands. This indicates that the strain and strain disorder present in the films induces modification in the hybridization between the Sr *4d* bands and the O *2p* states in films. Finally, there are also subtle variations in the metal-*sp* bands in both the films with respect to the bulk.

The XAS spectra provided important information about the spin state of the Ru ions. It is noted that the crystal field splitting is reduced in the films (it is lowest in S1 film). As a consequence, the spin state of the Ru ions may change. However, the intensity of the eg band is found to be the same in the two films and the bulk compound suggesting that the unoccupied states in the eg bands are same in all the three samples. In case of a spin state variation i.e. from low spin state in bulk to high spin state in films, the unoccupied density of state of the eg band in films should have shown a dramatic decrease. This should have reflected in decrease in intensity of the eg band in films which is contrary to the observation. Thus, it can be concluded that the films remain in the low spin state. The intensity of the t2g band on the other hand showed a systematic increase from bulk compound to S2 to S1. The increment in unoccupied density of states of the t2g can arise due to spin state variation or charge transfer. As mentioned above the possibility of high spin state is ruled out, thus charge transfer between metal ion (Ru t2g) and oxygen legend (O 2p) is the only remaining possibility. The effect of charge transfer is not merely reforming the nominal charges of the metal ions but complete change in the crystal field produced by the modified local oxygen environment as observed. It is well known that the charge transfer mechanism requires strong orbital overlapping. In order to probe the orbital overlap and the electronic structure near the fermi level, the valance band spectra was recorded. In figure 3, the valance band panel represents the resonant photoemission spectra at 52 eV for S1, S2 film and bulk. We observed that the position of the O *2p* bands is shifted towards the Ru *4d: t2g* band or the fermi level while the binding energy of the Ru *4d*: t2g band remains unaffected in the films when compared to the bulk. This clearly indicates relative change in the magnitude of the ligand-to-metal charge transfer energy Δ (the energy difference between the O *2p* band and Ru *4d*: t2g in the conduction band). It is found to be ~ 7 eV in the bulk compound while ~ 6 eV in the films. This shows that the O *2p* orbitals are strongly modified in thin films. The electronic structure is expected to get affected significantly due to the modifications in Ru *4d* and O *2p* orbital overlap. In order to probe the changes in electronic structure, we analysed the electronic structure using photoemission spectroscopy. The Ru *4p - 4d* resonant photoemission spectra of S1 and S2 films along with bulk SRO are measured in the on-resonance (52 eV) and the off-resonance (48 eV) condition at room temperature and is illustrated in figure 4 (a), (b), and (c), respectively. We followed the well-established electronic structure of SrRuO$_3$ in literature [25] and assigned the bands. Accordingly, the experimental features observed between ~ -4 eV and ~ -1 eV are assigned to the Ru *4d* band and from ~ -8 eV to ~ -3 eV are assigned to the O *2p*-derived band. A finite density of states of Ru *4d* band at the fermi level (E$_F$=0 eV) is observed in both the films and bulk SRO that is expected for a compound with good conductivity.

In the off-resonance condition, the spectrum is only due to the direct photoemission process:

$$Ru: 2p^6 4d^n + h\nu \rightarrow Ru: 2p^6 4d^{n-1} + e^-$$

While, in the on-resonance condition, the spectral features are enhanced due to intra-atomic excitation process or the indirect excitation process:

$$Ru: 2p^6 4d^n + h\nu \rightarrow [2p^5 3d^{n+1}]^* \rightarrow Ru: 2p^6 4d^{n-1} + e^-$$

We subtracted the valance band spectra recorded in on-resonance condition from the valance band spectra recorded in off-resonance condition, to deduce the direct contribution of the Ru *4d* derived states in the valance band. Such subtracted spectra contains positive and negative contributions as shown in the bottom panel of the figure 4. The positive contributions originate from the Ru *4d* component due to the resonant enhancement. On the other hand the O *2p* cross section decreases when going from $h\nu$=48 eV to $h\nu$=52 eV photon energy resulting in negative contributions of the derived states in the subtracted spectra. It transpires that the distribution of Ru derived states are different in S1, S2 films and Bulk. Figure 4 suggests that Ru derived states contribute prominently to the entire VBS in the S1 film, while its prominence is limited to the binding energy range of 1–7 eV for the S2 film and bulk. In other words, the distribution of Ru derived states is in descending order in S1, S2 and bulk SRO, respectively.

A careful observation of this figure suggests that the Ru *4d t2g* band is broader in S1 and S2 films when comparison to the bulk. The photoemission intensity around $E_F$ is also weaker in films than in the bulk. A direct visual evidences of these inferences are provided in the inset of figure 3 where the valance band spectra of the three samples near $E_F$ is projected in a zoomed condition. The broadening of the *d* band and decrease in the density of states at $E_F$ is considered as a hallmark of the presence of spin-orbit interactions [30,31].

The presence of spin-orbit interactions in film gives clue about the enhanced magnetic moment observed in films. As the spin state transition is ruled out, the only possibility for enhancement in magnetic moment is the contribution from the orbital moment. Generally, orbital moment is fully quenched in all the d-transition metal ions.

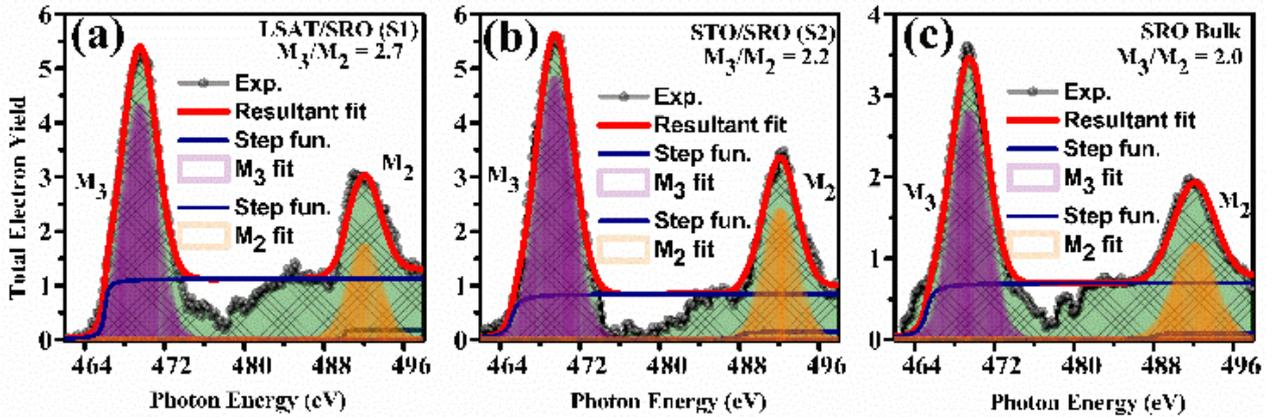

**Figure 5:-** The soft X-ray absorption spectroscopic spectra recorded on the LSAT/SRO (S1), STO/SRO (S2) and bulk SRO at the Ru ($M_{2,3}$-edges). The fitted spectra is also shown using a line.

However, when the spin-orbit coupling is strong, it disrupts the orbitals quenching. The signatures of the charge transfer observed in the XAS spectra collected at O K-edge also provides an evidence of the presence of spin-orbit coupling in the system. In order to estimate quantitatively, the spin-orbit coupling present in the system, XAS at the $M_{2,3}$ edges of the $Ru^{4+}$ ions is carried out and is presented in figure 6. Van der Laan and Thole [32,33, 34] showed that the intensity ratio of the white line features at $L_{2,3}$ or $M_{2,3}$ edges of the respective transition metal ion is directly proportional to the expectation value of the spin-orbit operator <L·S> = $\sum_i (l_i S_i)$. This method was successfully used in quantitative analysis of the spin-orbit coupling in iridium-based $5d$ compounds [35, 36]. The intensity ratio I(M$_3$)/I(M$_2$) =BR is defined as the branching ratio, where I(M$_{3,2}$) is the integrated intensity of the white line feature at the respective absorption edge [31,32, 37, 38]. The expectation value of the spin-orbit operator <L·S> is related with the branching ratio by the expression: BR = (2+r)/(1−r), where r = <L·S>/<n$_h$> and <n$_h$> refers to the average number of $4d$ holes. The value of <L·S> is expressed in units of $[(h/2\pi)^2]$. According to Van der Laan and Thole, a large <L·S> does imply the presence of strong spin-orbit coupling effects, however, the converse is not necessarily true. When the orbitals are fully quenched, the spin-orbit coupling strength is nearly zero giving the statistical average value of 2 for the branching ratio BR, while when the spin-orbit coupling is large the BR is expected to show higher values. Therefore, for the estimation of the spin-orbit coupling in the present samples, it is necessary to obtain accurate values of the integrated intensities of the white line features at the M$_2$ and M$_3$ absorption edges. For the same, the white line features were fitted using method described in ref. [31]. Accordingly, the white line was fitted by taking an arc tangent function for background contribution along with a Gaussian function for the white line feature. The fitting of the S1, S2 and bulk is presented in figure 5 (a), (b) and (c). The values thus obtained were used to calculate the BR which resulted in 2.7, 2.2 and 2.0 for S1, S2 and bulk SRO, respectively. It is noted that the value of the BR is significantly higher than the average statistical value of 2 in film S1 confirming presence of strong spin-orbit coupling. This explains the de-quenching of the orbitals resulting in enhanced magnetic moment in the S1 film.

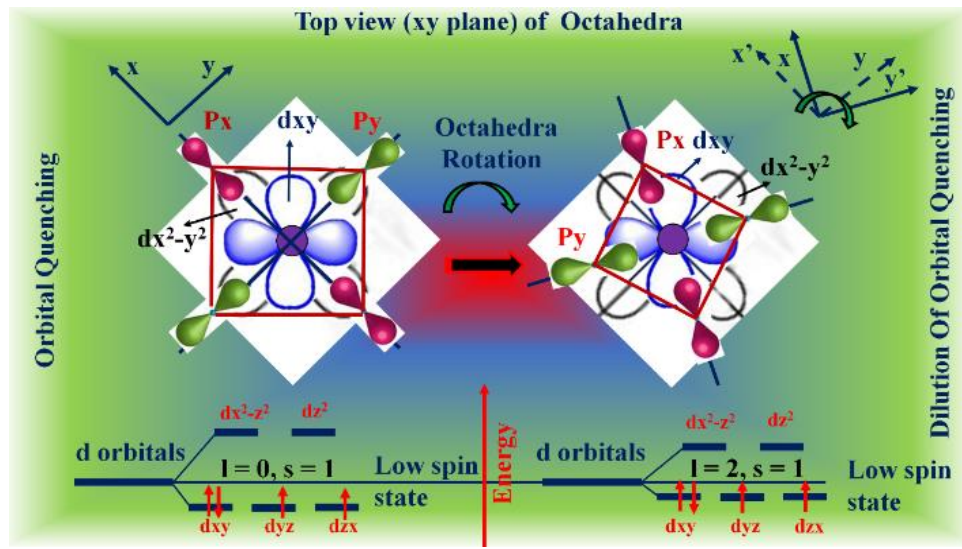

**Figure 6:** Schematic showing modifications in the hybridization of *Ru 4d* and *O 2p* orbitals due to strain disorder leading to disruption in the orbital quenching.

The mechanism responsible for the enhancement of the magnetic moment is presented in a schematic (figure 6). In bulk compound where the oxygen octahedral are arranged in a regular manner, the axial d-orbitals of the Ru ions ($dx^2$-$y^2$ and $dz^2$) are strongly hybridised with oxygen axial orbitals (*Px, Py and Pz*) leading to strong σ-bond (left panel of figure 6) which is responsible for the orbital quenching. On the other hand in films that possess large local strain inhomogeneity (as in S1 film in present case) the situation is different. In this case the oxygen octahedra are locally distorted in order to accommodate locally fluctuating strain state, leading to irregular tilts and rotations of the oxygen octahedra. This leads to reduction in the strength of the σ-bond and increase in the strength of the π-bond. This leads to partial de-quenching of the orbital moment. It is known that SrRuO$_3$ exhibits significant magneto-structural coupling and our study provides a strong evidence of it.

We have planned for detailed x-ray magnetic circular dichroism studies to further confirm this issue.

## 4. Conclusions:

Enhanced magnetic moment (3.3 $\mu_B$/Ru$^{4+}$) is demonstrated in the epitaxial film of SrRuO$_3$ grown on LSAT. In previous reports high spin state was attributed for the enhanced magnetic moment. In contrast, our results provide an evidence of low spin state and contribution of orbital angular moment responsible for the enhancement in the magnetic moment. As a rare experimental evidence the de-quenching of the orbital angular momentum is demonstrated in 4d transition metal oxide. The de-quenching of the orbitals introduces the strong spin-orbit coupling in this system. The strain disorder in the films is shown to be a tool to tune the hybridization between metal and ligand ions. The strain disorder also modifies other magnetic properties like coercivity apart from enhanced magnetic moment without disturbing the itinerant character of the films which have many applications.

**Acknowledgement:**

We would like to thank S.R. Barman for fruitful discussions and suggestions.